\documentstyle[12pt,graphicx]{article}

\newcommand{\loco}{{\mathop{ \, \rule[-.06in]{.2mm}{3.8mm}\,}}}

\newcommand{\doubar}{{{\loco}\!{\loco}}}

\def\mathfrak{\bf}

\def\be{\begin{equation}}
\def\ee{\end{equation}}
\def\bea{\begin{eqnarray}}
\def\eea{\end{eqnarray}}

\def\dt#1{\on{\hbox{\bf .}}{#1}}                
\def\Dot#1{\dt{#1}}
\def\IR{\relax{\rm I\kern-.18em R}}
\def\binomial#1#2{\left(\,{\buildrel
{\raise4pt\hbox{$\displaystyle{#1}$}}\over
{\raise-6pt\hbox{$\displaystyle{#2}$}}}\,\right)}

\def\[{\lfloor{\hskip 0.35pt}\!\!\!\lceil}
\def\]{\rfloor{\hskip 0.35pt}\!\!\!\rceil}

\newcommand{\AmS}{{\protect\the\textfont2
  A\kern-.1667em\lower.5ex\hbox{M}\kern-.125emS}}

\catcode`@=11
\def\un#1{\relax\ifmmode\@@underline#1\else
        $\@@underline{\hbox{#1}}$\relax\fi}
\catcode`@=12

\def\fracm#1#2{\hbox{\large{${\frac{{#1}}{{#2}}}$}}}

\def\ad{{\kern0.5pt
                   \alpha \kern-5.05pt
\raise5.8pt\hbox{$\textstyle.$}\kern
0.5pt}}

\def\Dot#1{{\kern0.5pt
     {#1} \kern-5.05pt \raise5.8pt\hbox{$\textstyle.$}\kern
0.5pt}}

\def\a{\alpha}
\def\b{\beta}

\def\d{\delta}
\def\e{\epsilon}

\def\g{\gamma}

\def\l{\lambda}
\def\m{\mu}

\def\o{\omega}

\def\s{\sigma}

\def\F{\Phi}

\def\O{\Omega}

\def\ca{{\cal A}}
\def\cb{{\cal B}}
\def\cc{{P}}

\def\ce{{\cal E}}
\def\cf{{\cal F}}

\def\cl{{\cal L}}

\def\co{{\cal O}}

\def\bo{{\raise.15ex\hbox{\large$\Box$}}}               
\def\TH{{\raise.2ex\hbox{$\displaystyle \bigodot$}\mskip-4.7mu \llap H
\;}}
\def\face{{\raise.2ex\hbox{$\displaystyle \bigodot$}\mskip-2.2mu \llap
{$\ddot
        \smile$}}}                                      

\def\Hat#1{\widehat{#1}}                        
\def\leftrightarrowfill{$\mathsurround=0pt \mathord\leftarrow \mkern-6mu
        \cleaders\hbox{$\mkern-2mu \mathord- \mkern-2mu$}\hfill
        \mkern-6mu \mathord\rightarrow$}
\def\dvec#1{\vbox{\ialign{##\crcr
        \leftrightarrowfill\crcr\noalign{\kern-1pt\nointerlineskip}
        $\hfil\displaystyle{#1}\hfil$\crcr}}}           
\def\dt#1{{\buildrel {\hbox{\LARGE .}} \over {#1}}}     

\def\fracm#1#2{\hbox{\large{${\frac{{#1}}{{#2}}}$}}}
\def\frac#1#2{{\textstyle{#1\over\vphantom2\smash{\raise.20ex
        \hbox{$\scriptstyle{#2}$}}}}}                   
\def\sfrac#1#2{{\vphantom1\smash{\lower.5ex\hbox{\small$#1$}}\over
        \vphantom1\smash{\raise.4ex\hbox{\small$#2$}}}} 
\def\bfrac#1#2{{\vphantom1\smash{\lower.5ex\hbox{$#1$}}\over
        \vphantom1\smash{\raise.3ex\hbox{$#2$}}}}       
\def\afrac#1#2{{\vphantom1\smash{\lower.5ex\hbox{$#1$}}\over#2}}    
\def\on#1#2{\mathop{\null#2}\limits^{#1}}               

\newskip\humongous \humongous=0pt plus 1000pt minus 1000pt
\def\caja{\mathsurround=0pt}
\def\eqalign#1{\,\vcenter{\openup2\jot \caja
        \ialign{\strut \hfil$\displaystyle{##}$&$
        \displaystyle{{}##}$\hfil\crcr#1\crcr}}\,}
\newif\ifdtup

  \def\pp{{\mathchoice
          {
              \kern 1pt%
              \raise 1pt
              \vbox{\hrule width5pt height0.4pt depth0pt
                    \kern -2pt
                    \hbox{\kern 2.3pt
                          \vrule width0.4pt height6pt depth0pt
                          }
                    \kern -2pt
                    \hrule width5pt height0.4pt depth0pt}%
                    \kern 1pt
           }
            {
              \kern 1pt%
              \raise 1pt
              \vbox{\hrule width4.3pt height0.4pt depth0pt
                    \kern -1.8pt
                    \hbox{\kern 1.95pt
                          \vrule width0.4pt height5.4pt depth0pt
                          }
                    \kern -1.8pt
                    \hrule width4.3pt height0.4pt depth0pt}%
                    \kern 1pt
            }
            {
              \kern 0.5pt%
              \raise 1pt
              \vbox{\hrule width4.0pt height0.3pt depth0pt
                    \kern -1.9pt  
                    \hbox{\kern 1.85pt
                          \vrule width0.3pt height5.7pt depth0pt
                          }
                    \kern -1.9pt
                    \hrule width4.0pt height0.3pt depth0pt}%
                    \kern 0.5pt
            }
            {
              \kern 0.5pt%
              \raise 1pt
              \vbox{\hrule width3.6pt height0.3pt depth0pt
                    \kern -1.5pt
                    \hbox{\kern 1.65pt
                          \vrule width0.3pt height4.5pt depth0pt
                          }
                    \kern -1.5pt
                    \hrule width3.6pt height0.3pt depth0pt}%
                    \kern 0.5pt
            }
        }}

  \def\mm{{\mathchoice
                       {
                             \kern 1pt
               \raise 1pt    \vbox{\hrule width5pt height0.4pt depth0pt
                                  \kern 2pt
                                  \hrule width5pt height0.4pt depth0pt}
                             \kern 1pt}
                       {
                            \kern 1pt
               \raise 1pt \vbox{\hrule width4.3pt height0.4pt depth0pt
                                  \kern 1.8pt
                                  \hrule width4.3pt height0.4pt depth0pt}
                             \kern 1pt}
                       {
                            \kern 0.5pt
               \raise 1pt
                            \vbox{\hrule width4.0pt height0.3pt depth0pt
                                  \kern 1.9pt
                                  \hrule width4.0pt height0.3pt depth0pt}
                            \kern 1pt}
                       {
                           \kern 0.5pt
             \raise 1pt  \vbox{\hrule width3.6pt height0.3pt depth0pt
                                  \kern 1.5pt
                                  \hrule width3.6pt height0.3pt depth0pt}
                           \kern 0.5pt}
                       }}

\def\pd{{\kern0.5pt
                   + \kern-5.05pt \raise5.8pt\hbox{$\textstyle.$}\kern
0.5pt}}

\def\pmd{{\kern0.5pt
                  \pm \kern-5.05pt \raise6.3pt\hbox{$\textstyle.$}\kern1.5pt}}

\def\md{{\mathchoice
   {
      {{\kern 1pt - \kern-6.2pt \raise5pt\hbox{$\textstyle.$}\kern 1pt}}}
    {
      {{\kern 1pt - \kern-6.2pt \raise5pt\hbox{$\textstyle.$}\kern 1pt}}}
    {
      {\kern0.5pt - \kern-5.05pt \raise3.4pt\hbox{$\textstyle.$}\kern0.5pt}}
    {
      {\kern0.5pt - \kern-5.05pt \raise3.4pt\hbox{$\textstyle.$}\kern0.5pt}}}}

\def\ad{{\dot{\alpha}}}

\def\pp{{\mathchoice
          {
              \kern 1pt%
              \raise 1pt
              \vbox{\hrule width5pt height0.4pt depth0pt
                    \kern -2pt
                    \hbox{\kern 2.3pt
                          \vrule width0.4pt height6pt depth0pt
                          }
                    \kern -2pt
                    \hrule width5pt height0.4pt depth0pt}%
                    \kern 1pt
           }
            {
              \kern 1pt%
              \raise 1pt
              \vbox{\hrule width4.3pt height0.4pt depth0pt
                    \kern -1.8pt
                    \hbox{\kern 1.95pt
                          \vrule width0.4pt height5.4pt depth0pt
                          }
                    \kern -1.8pt
                    \hrule width4.3pt height0.4pt depth0pt}%
                    \kern 1pt
            }
            {
              \kern 0.5pt%
              \raise 1pt
              \vbox{\hrule width4.0pt height0.3pt depth0pt
                    \kern -1.9pt  
                    \hbox{\kern 1.85pt
                          \vrule width0.3pt height5.7pt depth0pt
                          }
                    \kern -1.9pt
                    \hrule width4.0pt height0.3pt depth0pt}%
                    \kern 0.5pt
            }
            {
              \kern 0.5pt%
              \raise 1pt
              \vbox{\hrule width3.6pt height0.3pt depth0pt
                    \kern -1.5pt
                    \hbox{\kern 1.65pt
                          \vrule width0.3pt height4.5pt depth0pt
                          }
                    \kern -1.5pt
                    \hrule width3.6pt height0.3pt depth0pt}%
                    \kern 0.5pt
            }
        }}

  \def\mm{{\mathchoice
                       {
                             \kern 1pt
               \raise 1pt    \vbox{\hrule width5pt height0.4pt depth0pt
                                  \kern 2pt
                                  \hrule width5pt height0.4pt depth0pt}
                             \kern 1pt}
                       {
                            \kern 1pt
               \raise 1pt \vbox{\hrule width4.3pt height0.4pt depth0pt
                                  \kern 1.8pt
                                  \hrule width4.3pt height0.4pt depth0pt}
                             \kern 1pt}
                       {
                            \kern 0.5pt
               \raise 1pt
                            \vbox{\hrule width4.0pt height0.3pt depth0pt
                                  \kern 1.9pt
                                  \hrule width4.0pt height0.3pt depth0pt}
                            \kern 1pt}
                       {
                           \kern 0.5pt
             \raise 1pt  \vbox{\hrule width3.6pt height0.3pt depth0pt
                                  \kern 1.5pt
                                  \hrule width3.6pt height0.3pt depth0pt}
                           \kern 0.5pt}
                       }}

\def\pd{{\kern0.5pt
                   + \kern-5.05pt \raise5.8pt\hbox{$\textstyle.$}\kern
0.5pt}}

\def\pmd{{\kern0.5pt
                  \pm \kern-5.05pt \raise6.3pt\hbox{$\textstyle.$}\kern1.5pt}}

\def\md{{\mathchoice
   {
      {{\kern 1pt - \kern-6.2pt \raise5pt\hbox{$\textstyle.$}\kern 1pt}}}
    {
      {{\kern 1pt - \kern-6.2pt \raise5pt\hbox{$\textstyle.$}\kern 1pt}}}
    {
      {\kern0.5pt - \kern-5.05pt \raise3.4pt\hbox{$\textstyle.$}\kern0.5pt}}
    {
      {\kern0.5pt - \kern-5.05pt \raise3.4pt\hbox{$\textstyle.$}\kern0.5pt}}}}

\def\dslash{\not{\hbox{\kern-2pt $\partial$}}}
\def\Dslash{\not{\hbox{\kern-4pt $D$}}}
\def\pslash{\not{\hbox{\kern-2.3pt $p$}}}
 \newtoks\slashfraction
 \slashfraction={.13}
 \def\slash#1{\setbox0\hbox{$ #1 $}
 \setbox0\hbox to \the\slashfraction\wd0{\hss \box0}/\box0 }

\font\ro=cmsy10                          
\def\kcr{{\hbox{\ro \char'170}}}                
\def\ktl{{\hbox{\ro \char'170}}}        
\def\ktr{{\hbox{\ro \char'170}}}        
\def\kbl{{\hbox{\ro \char'170}}}        
\def\kbr{{\hbox{\ro \char'170}}}        

\def\plpl{\raise-2pt\hbox{$\raise3pt\hbox{$_+$}\hskip-6.67pt\raise0.0pt
\hbox{$^+$}\hskip 0.01pt$}}
\def\mimi{\raise-2pt\hbox{$\raise3pt\hbox{$_-$}\hskip-6.67pt\raise0.0pt
\hbox{$^-$}\hskip 0.01pt$}}

\def\bo{{\raise.15ex\hbox{\large$\Box$}}}               
\def\TH{{\raise.2ex\hbox{$\displaystyle \bigodot$}\mskip-4.7mu \llap H \;}}
\def\face{{\raise.2ex\hbox{$\displaystyle \bigodot$}\mskip-2.2mu \llap {$\ddot
        \smile$}}}                                      

\def\Hat#1{\widehat{#1}}                        
\def\leftrightarrowfill{$\mathsurround=0pt \mathord\leftarrow \mkern-6mu
        \cleaders\hbox{$\mkern-2mu \mathord- \mkern-2mu$}\hfill
        \mkern-6mu \mathord\rightarrow$}
\def\dvec#1{\vbox{\ialign{##\crcr
        \leftrightarrowfill\crcr\noalign{\kern-1pt\nointerlineskip}
        $\hfil\displaystyle{#1}\hfil$\crcr}}}           
\def\dt#1{{\buildrel {\hbox{\LARGE .}} \over {#1}}}     

\def\fracm#1#2{\hbox{\large{${\frac{{#1}}{{#2}}}$}}}
\def\frac#1#2{{\textstyle{#1\over\vphantom2\smash{\raise.20ex
        \hbox{$\scriptstyle{#2}$}}}}}                   
\def\sfrac#1#2{{\vphantom1\smash{\lower.5ex\hbox{\small$#1$}}\over
        \vphantom1\smash{\raise.4ex\hbox{\small$#2$}}}} 
\def\bfrac#1#2{{\vphantom1\smash{\lower.5ex\hbox{$#1$}}\over
        \vphantom1\smash{\raise.3ex\hbox{$#2$}}}}       
\def\afrac#1#2{{\vphantom1\smash{\lower.5ex\hbox{$#1$}}\over#2}}    
\def\on#1#2{\mathop{\null#2}\limits^{#1}}               

\parskip=\medskipamount                 
\thicklines                         

\def\oldheadpic{                                
        \setlength{\unitlength}{.4mm}
        \thinlines
        \par
        \begin{picture}(349,16)
        \put(325,16){\line(1,0){4}}
        \put(330,16){\line(1,0){4}}
        \put(340,16){\line(1,0){4}}
        \put(335,0){\line(1,0){4}}
        \put(340,0){\line(1,0){4}}
        \put(345,0){\line(1,0){4}}
        \put(329,0){\line(0,1){16}}
        \put(330,0){\line(0,1){16}}
        \put(339,0){\line(0,1){16}}
        \put(340,0){\line(0,1){16}}
        \put(344,0){\line(0,1){16}}
        \put(345,0){\line(0,1){16}}
        \put(329,16){\oval(8,32)[bl]}
        \put(330,16){\oval(8,32)[br]}
        \put(339,0){\oval(8,32)[tl]}
        \put(345,0){\oval(8,32)[tr]}
        \end{picture}
        \par
        \thicklines
        \vskip.2in}
\def\oldtitle#1#2#3#4{\oldheadpic\begin{center}\vglue.5in{\large\bf #1}\\[.6in]
        {#2}\\[.1in] {\it Department of Physics and Astronomy}\\
        {\it University of Maryland, College Park, MD 20742}\\[.6in]
        Physics Publication \#{#3}\\ {#4}\\[1.5in] {\bf ABSTRACT}\\[.1in]
        \end{center} \begin{quotation}}                 
\def\oldTitle#1#2#3#4#5#6#7{\oldheadpic\begin{center} \vglue .4in
        {\large\bf #1}\\[.4in]
        {#2}\\[.1in] {\it Department of Physics and Astronomy}\\
        {\it University of Maryland, College Park, MD 20742}\\[.1in]
        {#3}\\[.1in] {\it {#4}}\\ {\it {#5}}\\[.4in]
        Physics Publication \#{#6}\\ {#7}\\[.5in] {\bf ABSTRACT}\\[.1in]
        \end{center} \begin{quotation}}                 
\def\border{                                            
        \setlength{\unitlength}{1mm}
        \newcount\xco
        \newcount\yco
        \xco=-21
        \yco=12
        \begin{picture}(140,0)
        \put(\xco,\yco){$\ktl$}
        \advance\yco by-1
        {\loop
        \put(\xco,\yco){$\kcr$}
        \advance\yco by-2
        \ifnum\yco>-240
        \repeat
        \put(\xco,\yco){$\kbl$}}
        \xco=158
        \yco=12
        \put(\xco,\yco){$\ktr$}
        \advance\yco by-1
        {\loop
        \put(\xco,\yco){$\kcr$}
        \advance\yco by-2
        \ifnum\yco>-240
        \repeat
        \put(\xco,\yco){$\kbr$}}
        \put(-20,13){\tiny **University of Maryland * Center for String and
         Particle  Theory* Physics Department***University of Maryland *Center
        for String and Particle  Theory** }
        \put(-20,-241.5){\tiny **University of Maryland * Center for String and
         Particle  Theory* Physics Department***University of Maryland *Center
        for String and Particle  Theory** }
        \end{picture}
        \par\vskip-8mm}
\def\bordero{                                           
        \setlength{\unitlength}{1mm}
        \newcount\xco
        \newcount\yco
        \xco=-31
        \yco=12
        \begin{picture}(140,0)
        \put(\xco,\yco){$\ktl$}
        \advance\yco by-1
        {\loop
        \put(\xco,\yco){$\kclr}
        \advance\yco by-2
        \ifnum\yco>-240
        \repeat
        \put(\xco,\yco){$\kbl$}}
        \xco=151
        \yco=12
        \put(\xco,\yco){$\ktr$}
        \advance\yco by-1
        {\loop
        \put(\xco,\yco){$\kcr$}
        \advance\yco by-2
        \ifnum\yco>-240
        \repeat
        \put(\xco,\yco){$\kbr$}}
        \put(-20,12){\ooo bacdefghidfghghdhededbihdgdfdfhhdheidhdhebaaahjhhdahba

hgdedge
   hgfdiehhgdigicba}
        \put(-20,-241.5){\ooo ababaighefdbfghgeahgdfgafagihdidihiidhiagfedhadbfd

ecdcdfa
   gdcbhaddhbgfchbgfdacfediacbabab}
        \end{picture}
        \par\vskip-8mm}
\def\headpic{                                           
        \indent
        \setlength{\unitlength}{.4mm}
        \thinlines
        \par
        \begin{picture}(29,16)
        \put(165,16){\line(1,0){4}}
        \put(170,16){\line(1,0){4}}
        \put(180,16){\line(1,0){4}}
        \put(175,0){\line(1,0){4}}
        \put(180,0){\line(1,0){4}}
        \put(185,0){\line(1,0){4}}
        \put(169,0){\line(0,1){16}}
        \put(170,0){\line(0,1){16}}
        \put(179,0){\line(0,1){16}}
        \put(180,0){\line(0,1){16}}
        \put(184,0){\line(0,1){16}}
        \put(185,0){\line(0,1){16}}
        \put(169,16){\oval(8,32)[bl]}
        \put(170,16){\oval(8,32)[br]}
        \put(179,0){\oval(8,32)[tl]}
        \put(185,0){\oval(8,32)[tr]}
        \end{picture}
        \par\vskip-6.5mm
        \thicklines}
\def\title#1#2#3#4{\border\headpic {\hbox to\hsize{#4 \hfill UMDEPP #3}}\par
        \begin{center} \vglue .5in {\large\bf #1}\\[.6in]
        {#2}\\[.1in] {\it Department of Physics and Astronomy}\\
        {\it University of Maryland, College Park, MD 20742}\\[1.5in]
        {\bf ABSTRACT}\\[.1in] \end{center} \begin{quotation}}  
\def\Title#1#2#3#4#5#6#7{\border\headpic
        {\hbox to\hsize{#7 \hfill UMDEPP #6}}\par
        \begin{center} \vglue .4in {\large\bf #1}\\[.4in]
        {#2}\\[.1in] {\it Department of Physics and Astronomy}\\
        {\it University of Maryland, College Park, MD 20742}\\[.1in]
        {#3}\\[.1in] {\it {#4}}\\ {\it {#5}}\\[.5in] {\bf ABSTRACT}\\[.1in]
        \end{center} \begin{quotation}}                 
\def\endtitle{\end{quotation}\newpage}                  

\def\qd{{\kern0.5pt
                   q \kern-5.05pt \raise5.8pt\hbox{$\textstyle.$}\kern
0.5pt}}

\begin{document}

\def\dt#1{\on{\hbox{\bf .}}{#1}}                
\def\Dot#1{\dt{#1}}

\def\gfrac#1#2{\frac {\scriptstyle{#1}}
        {\mbox{\raisebox{-.6ex}{$\scriptstyle{#2}$}}}}
\def\gg{{\hbox{\sc g}}}
\border\headpic {\hbox to\hsize{September 2004 \hfill
{UMDEPP 05-010}}}
\par
{$~$ \hfill
{hep-th/0409104}}
\par

\setlength{\oddsidemargin}{0.3in}
\setlength{\evensidemargin}{-0.3in}
\begin{center}
\vglue .10in
{\large\bf Dynamical Equations from a First-Order\\
Perturbative Superspace Formulation of\\[.1in]
10D, $\cal N$ = 1 String-Corrected Supergravity
(I)\footnote
{Supported in part  by National Science Foundation Grant
PHY-0354401.}\  }
\\[.5in]

S.\, James Gates, Jr.\footnote{gatess@wam.umd.edu},
Annam\' aria Kiss\footnote{akiss@physics.umd.edu}
and Willie Merrell\footnote{williem@physics.umd.edu}
\\[0.2in]

{\it Center for String and Particle Theory\\
Department of Physics, University of Maryland\\
College Park, MD 20742-4111 USA}\\[1.5in]

{\bf ABSTRACT}\\[.01in]
\end{center}
\begin{quotation}
{Utilizing a first-order perturbative superspace approach, we
derive the bosonic equations of motion for the 10D, $\cal N$ = 1
supergravity fields. We give the Lagrangian corresponding to these
equations derived from superspace geometry. Moreover, the
equivalence of this Lagrangian to the first-order perturbative
component level Lagrangian of anomaly-free supergravity is proven.
Our treatment covers both the two-form and six-form formulations.}

${~~~}$ \newline

\endtitle

\section{Introduction}

~~~~ Some years ago \cite{10DalphPrm,10DTblisi,10Defx,10DCSform,
10DCSform2}, we began to set up a framework for development of a perturbative
superspace method to describe the effect of string corrections to the
equations of motion for 10D, $\cal N$ = 1 supergravity.  At that time, we
emphasized the importance of a perturbative approach as being best
suited to such an application.  Today such perturbative modifications to
superspace geometry are often called ``deformations.''

A little prior to the appearance of the work in \cite{10DalphPrm},
there was a {\em {suggestion}} \cite{AtkDhRTRA} that it should be
possible to use superspace Bianchi identities to find the string
corrected equations of motion for 10D, $\cal N$ = 1 supergravity.
However, this work {\em {contained no}} concrete or clear
indication of how this might be done. After the appearance of
\cite{10DalphPrm} Nilsson \cite{NLEEA} performed an analysis of
the usual superspace geometry, based on the constraint $T_{\a
\, \b}{\,}^{\un c} ~=~ i \, (\s^{\un c})_{\a \, \b }$ (in our
notation), and concluded that it requires modification in the
presence of string corrections.

Other groups (see references [3-13] in the work of \cite{IT})
during this period began to study this issue also.  Two notable
features of this class of works are: (a.) the approach is
described as being a {\em {non-perturbative}} description (i.e.
contains string corrections correctly to all orders) and (b.) the
condition $T_{\a \, \b}{\,}^{\un c} ~=~ i \, (\s^{\un c})_{\a \,
\b } $ on the lowest dimensional torsion component is imposed to
{\em {all}} orders in the string corrections.

The above recitation of the published historical record is thus completely
clear in determining that discussion of the perturbative deformation technique
begins with the work in \cite{10DalphPrm} despite impressions otherwise that
may be engendered by many patterns of citation on this topic in today's literature.
One of the authors of this work (SJG) claims, even at the present point in time, that
the deformations identified in \cite{10DCSform,10DCSform2} provide a correct
{\em {first-order}} description of the Lorentz Chern-Simons modifications to
10D, $\cal N$ = 1 supergravity and has never accepted the argument given
in \cite{IT}.  The reasons for this will be given in an accompanying paper
\cite{Pap2}.

In an attempt to initiate a logical and calm debate with a minimal
of controversy, in this work we investigate {\em {solely}} the
proposal of \cite{ 10DCSform,10DCSform2}.
Namely, we concentrate only on an investigation at first order in
the string tension.  We undertake this not to re-ignite what
became an acrimonious debate but because more recent developments
suggest this is the appropriate time to re-enter this debate.

At the time we first began deliberations of how closed type-I and heterotic string
theory must perturbatively modify superspace geometry, we also similarly investigated
how the open type-I superstring must have a similar effect on 10D, $\cal N$ = 1 super
Yang-Mills theory.  This effort was rewarded with the discovery of the first
manifestly supersymmetrical description (see first work of \cite{10Defx}) of the
lowest order correction from open superstring theory.  Again the method used
was that of a perturbative solution to the appropriate superspace Bianchi
identities.  The component formulation of the lowest order pure Yang-Mills
part of the corrections was discovered in \cite{Tsey}. Very shortly after our
superspace description appeared in \cite{10Defx}, an equivalent component
level derivation was carried out \cite{BRakSzg} and complete agreement was
found.  However, neither the result in the first work of \cite{10Defx} nor that
in  \cite{BRakSzg} was derived {\em {directly}} from a superstring argument.
This same result was verified yet again more recently \cite{KER} and more
importantly a recursive procedure was developed to determine the explicit
form of the deformation to {\em {all}} orders.  But again the method used did
not rely on superstring theory.

Finally in 2002, the first derivation \cite{BERK} of the lowest
order open string correction to superspace deformations (along
with a recursive procedure to derive all orders results) was found
on the basis of superstring theory.  Complete agreement was found
with the result in {\cite{10Defx}}.  As the reader will note this
is eighteen years after our first suggestion of this lowest order
superspace deformation arising from the open superstring.   This
gulf of time is indicative of the difficulty and subtlety of
direct superstring derivations of superspace descriptions of the
string low energy effective action.

The ``pure spinor'' method used in \cite{BERK} is clearly a very powerful tool
for providing superstring-based derivations of the required deformations
to supergeometry.  With continued development (e.g. \cite{Grassi}), we expect
that at some point in the not too distant  future it will be applied to the problem
of the closed type-I or heterotic superstring low energy effective action.

In light of this, we wish to have in the record (and in advance of
this hoped-for breakthrough using the pure spinor technique in
covariant superstring field theory) as complete as possible a
description of the proposal for the superspace supergravity
deformations made in the works of \cite{10DCSform,10DCSform2}.
This is the primary reason for our offering the present work for
the consideration of our readers.  A pure spinor derivation of the
deformation should definitively settle the controversy.

Our conventions and definitions for 10D metrics, Pauli matrices
have appeared in many of the works in \cite{10DalphPrm} -
\cite{10DCSform2} and we have also included an additional
discussion of definitions in an appendix. A mathematica package
for manipulating our 10D spinor matrix algebra is available
on-line via hep-th/0004202 \cite{PACKG}.

\section{A Review of First-Order Corrected 10D, $\cal N$ = 1 Superspace
Supergravity Geometry}

~~~~ Let us begin by reviewing the results in \cite{10DCSform2}.
There a solution was given to the 10D, $\cal N$ = 1 supergravity
Bianchi identities that is correct to first order in the
perturbative parameters $\b^\prime$ and $\g'$. Here we recall the
Yang-Mills truncated version (eliminating the Yang-Mills fields or
equivalently putting $\b^\prime=0$) of this solution and we
complete the results by the presentation of the bosonic equations
of motion. Also, we would like to emphasize here that this
solution rests on two assumptions, both valid at first order in
$\g^\prime$. The first assumption is a constraint on the 0
dimensional torsion component $T_{\d\g}{}^{\un a}$ and the second
is a choice for the usual 1 dimensional auxiliary field denoted by
$A_{\un a\,\un b\,\un c}$. Let us now discuss these two inputs and
their consequences in order.

Before making any assumptions on torsion components it is always
worthwhile to study purely conventional constraints, which do
reduce the number of independent torsion and curvature components,
but do not have any consequence on the dynamics. Using standard
methods (see the work in \cite{11Dconstr}) one can show that the
following set of constraints is purely conventional:
\be
\eqalign{ ~ &i \, (\s_{\un a})^{\a \b} \, T_{\a \b}{ }^{\un b} ~=~
16 \, \d_{\un a} \, {}^{\un b}  ~~,~~
 i \, (\s_{\un c})^{\a \b} \, T_{\a \, \un b}{}^{\un c} ~=~ 0 ~\,~,
   \label{eq:Rev2.4}  \cr
&i \, (\s_{\un a \, \un b \, \un c \, \un d \,  \un e})^{\a \b} \,
T_{\a \b}{ }^{\un e} ~=~ 0 ~~~~,~~ T_{\a \, [ \un d \, \un e]}
~=~0 ~~~~~~~~\,~,
   \label{eq:Rev2.5}  \cr
&T_{\un d \, \un e \, \un b} ~=~ \fracm18 \, (\s_{\un d \, \un e
}){}_{\a}{}^{\b} \,  T_{\b \, \un b}{}^{\a} ~+~ i \, \fracm1{16}
\, (\s_{\un b})^{\a \,\b} \, {\cal R}_{\a \, \b}{\,}_{\un d \, \un
e}  ~\,~. }     \label{eq:Rev2.6}
\label{EQ:constr}
\ee
The role of each of these respective constraints is easy to
understand. The first equation removes E${}_{\un a}{}^{\un m}$ as
an independent variable.  The second equation removes E${}_{\un
a}{}^{\m}$ as an independent variable.  The third constraint is a
coset conventional constraint that removes part of
E${}_{\a}{}^{\m}$ as an independent variable.  The fourth
constraint removes $\o{}_{\a \, \un b \, \un c}$ as  an
independent variable and the final constraint removes $\o{}_{\un a
\, \un b \, \un c}$ as  an independent variable.  It is a simple
matter to show that the torsion and curvature super tensors in
\cite{10DCSform2}, satisfy these conditions.  Since these are
purely conventional constraints, they may be imposed to {\em {all}}
orders in the string slope-parameter expansion.

In addition, notice that the first and third equations of
(\ref{EQ:constr}) imply that the most general structure of the
zero dimensional torsion is
\be
T_{\d\g}{}^{\un a} \ =\ i \,(\s^{\un a})_{\d\g} ~+~  i \, \fracm
1{5!} \,  (\s^{[5]} )_{\d\g} \, X_{[5]}{}^{\un a}  ~~~,
\label{eq:Rev2.2}
\ee
with $X_{[5]}{}^{\un a}$ in the appropriate 1050 dimensional irrep
of $SO(1,9)$, while in \cite{10DCSform2} the following zero
dimensional torsion constraint was {\em {assumed}}
\be
T_{\d\g}{}^{\un a} \, =\, i \,(\s^{\un a})_{\d\g} + {\cal O}
{\Huge (} \,    (\g')^2    \,   {\Huge )}  ~~~.
\label{eq:Rev2.1}
\ee
Recall also, that  Nilsson advocated already in \cite{NLEEA} that the
assumption $X_{[5]}{}^{\un a}=0$ is incompatible with the
inclusion of higher than second order curvature terms in the
effective action. In this logic the vanishing of this 0
dimensional superfield can be valid only at first order in
$\g^\prime$ -- the regime in which \cite{10DCSform2} was written.
It is worthwhile to note the possibility that $X_{[5]}{}^{\un a}$ contributes to
higher order terms, \be \eqalign{
X_{[5]}{}^{\un a}  ~=~ {\cal O} {\Huge (} \,    (\g')^2    \,
{\Huge )}  ~~~. }    \label{eq:Rev2.3}
\ee
 But as the work in \cite{10DCSform2} was only to
first order, it was completely moot on this point.

Now let us begin to write the solution of the Bianchi identities
for the torsion subject to the conventional constraints
(\ref{EQ:constr}) and assumption (\ref{eq:Rev2.1}):
\bea
T_{\a\b}{}^\g&=&-\left[ ~ \d^\g_{(\a}\d^\d_{\b)}\,+\, (\s^{\un a})_{\a\b}(\s_{
\un a})^{\g\d} ~
\right]\chi_\d   ~~~,    \label{eq:Rev2.7}  \\
T_{\a \un b}{}^\g&=&\frac{1}{48}(\s_{\un b} \, \s^{[3]})_\a{}^\g A_{[3]}
~~~,     \label{eq:Rev2.8}  \\
T_{\un a \, \un b \, \un c}&=&-2 \, L_{\un a \, \un b \, \un c}
~~~,      \label{eq:Rev2.9} \\[2mm]
{\cal R}_{\a\, \b ~ \un a \, \un b }&=&i \,2 (\s^{\un c})_{\a\b}( ~ L_{\un a \, \un b \, \un c} \,
-\, \frac{1}{8}A_{\un a \, \un b \, \un c} \, \,) ~-~
i \, \frac{1}{24}(\s_{\un a \, \un b \, \un c \,  \un d \,  \un e})_{\a\b}A^{
 \un c \,  \un d \,  \un e} ~~~,    \label{eq:Rev2.10}  \\[2mm]
\nabla_\a\chi_\b&=&-i \,(\s^{\un a})_{\a\b} \nabla_{\un a} \Phi
~+~ i \, \frac{1}{48}(\s^{[3]})_{\a\b}\left( \, 4L_{[3]} \,+\, A_{[3]} \,-\,
i \frac{1}{2}(\chi\s_{[3]}\chi)  \, \right) ~~~,    \label{eq:Rev2.11}  \\[2mm]
{\cal R}_{\a \un c \, \un a \, \un b}&=&i \,(\s_{[\un
a|})_{\a\b}T_{\un c| \, \un b]}{}^\b + i \g^\prime\,(\s_{[\un
c|})_{\a\b}{\cal R}^{\un k \, \un l}{}_{|\un a \,
\un b]}T_{\un k \, \un l}{}^\b ~~~,
 \label{cdki}
 \label{eq:Rev2.12}
\eea
with $\Phi$ a scalar superfield (dilaton) transforming into $\chi_\a$ (dilatino)
under supersymmetry,
\be
\chi_\a\ =\ -2\, \nabla_\a \Phi ~~~,
 \label{eq:Rev2.13}
\ee
and $A_{\un a \, \un b \, \un c}$ an auxiliary superfield.  Once again, we emphasize
these are required {\em {solely}} to first order in $\g^{\prime}$.

The fixing of this auxiliary field is the second input which
characterizes the solution proposed in \cite{10DCSform2}. Making
the {\it choice}
\bea
A_{\un a \, \un b \, \un c}&\doteq&-i \,\g^\prime  \, (T_{\un k \, \un l} \, \s_{
\un a \, \un b \, \un c} \, T^{\un k \, \un l}) ~~~,
  \label{eq:Rev2.14}
\eea
the theory is put completely on shell. This means that all torsion
and curvature components, as well as the spinorial derivatives of
all objects in the geometry can be expressed in function of the
dilaton $\Phi$, the dilatino $\chi_\a$, the gravitino Weyl tensor
sitting in its field strengths $T_{\un a \, \un b}{}^\g$,  the
Weyl tensor sitting in the curvature ${\cal R}_{\un a \, \un b \,
\un c \, \un d}$ together with  the supercovariant object $L_{\un
a \, \un b \, \un c}$ appearing in the spacetime torsion.

The form of $A_{\un a \, \un b \, \un c}$ written in terms of
$T_{\un a  \un b} {}^{\g}$ (the supercovariantized gravitino
``curl'') can be understood on the basis of a remarkable property:
open-string/closed-string duality. This conjectured property of
the low energy effective action of the superstring was made even
before this property had a name.  Bergshoeff and Rakowski
\cite{BergRak} noted that in 6D simple superspace the
quantities
\be \eqalign{
 T^{\un c \, \un d}{\,}^{\g} ~~,~~ {\cal R}_{\un a \, \un b}{\,}^{ \, \un c  \, \un d}
}    \label{eq:Rev2.15}
\ee
share many common properties with the fields of
a vector multiplet
\be \eqalign{
 \l^{\g \, {\Hat I}} ~~,~~ F_{\un a \, \un b}{\,}^{ {\Hat I}}
}    \label{eq:Rev2.16}
\ee
and thus asserted that large numbers of higher derivative
supergravity terms may be treated as if one were coupling a vector
multiplet to the supergavity multiplet.  The result in \cite{10DCSform2}
implements this strategy for the 10D, ${\cal N} \, = ~1$ superspace
and after this work Bergshoeff and de Roo \cite{BergdROO} extended
this approach to 10D theories at the level of component fields.

The object $L_{\un a \, \un b \, \un c}$ was introduced\footnote{The first
appearance of the $L$-type variable in the physics literature occurred
in the work \newline $~~~~~~$ of \cite{Lref}. It was introduced to permit
a unified superspace description of theories related \newline $~~~~~~$
 one to another by Poincar\' e duality.} for the ten dimensional theory
\cite{10DCSform2} in order to permit the simple passage between the
2-form and 6-form formulation of the 10D, $\cal N$ $= ~1$  supergravity
theory. It is not an independent variable but its explicit form as
a function of the component fields is determined only by
specifying which of the two (2-form vs.\ 6-form) gauge fields is in
the supergravity multiplet. This will be discussed in subsequent
chapters.

In particular, $L_{\un a \, \un b \, \un c}$ must satisfy the
following conditions
\bea
\nabla_\a L_{\un a \, \un b \, \un c }&=&i \, \frac{1}{4}(\s_{[\un a
})_{\a\b} \, ( \, \, T_{\un b \, \un c ]}{}^\b \,-\, \g^\prime {\cal R}^{\un k \, \un l}{}_{
\un b \, \un c ]} \, T_{\un k \un l}{}^\b ~) ~~~,    \label{eq:Rev2.17} \\
\nabla_\a T_{\un a \, \un b }{}^\b&=&\frac{1}{4}(\s^{\un c \, \un d})_\a
{}^\b \, {\cal R}_{\un a \, \un b \, \un c \, \un d} \,-\, T_{\un a \, \un b}{}^\g \,
T_{\g\a}{}^\b   \nonumber \\
&&+\frac{1}{48}\left[~ 2L_{\un a \, \un b \, \un c}(\s^{\un c} \, \s^{[3]})_\a{}^\b
\,-\, (\s_{[\un a|}\s^{[3]})_\a{}^\b\nabla_{|\un b]} \, \, \right] \, A_{[3]} ~~~,
\label{eq:Rev2.18}
\eea
in order for the Bianchi identities on the superspace torsions and curvatures to
be satisfied.  These same Bianchi identities require
\bea
\nabla_{\un a} \chi_\b&=&-i \, \frac{1}{2}(\s^{\un b})_{\a\b}(T_{\un a \, \un b}{}^\a
\,-\, 2\g^\prime {\cal R}^{\un k \, \un l}{}_{\un a \un b}  \, T_{\un k \, \un l}{}^\a)\label{Dki}
~~~,  \label{eq:Rev2.19}~~~ \\
(\s^{\un a \, \un b})_\b{}^\a T_{\un a \, \un b}{}^\b&=&-i \,8 \,(\s^{\un a})^{\a\b}\chi_\b
\nabla_{\un a}  \Phi \,-\, i
\frac{1}{24}(\s^{[3]})^{\a\b}\chi_\b \,( ~ 16L_{[3]} \,+\, A_{[3]} ~) \nonumber \\
&&+\, 3 \g^\prime(\s^{\un a \, \un b})_\b{}^\a {\cal R}^{\un k \, \un l}{}_{\un a \, \un b}
\, T_{\un k \, \un l}{}^\b ~~~.
\label{eq:Rev2.20}
\eea
The results given above are sufficient to derive the equations of
motion for the spinors, already presented in \cite{10DCSform2},
and we will now use them in order to derive the bosonic equations
of motion. A detailed presentation of using superspace techniques
for deriving equations of motion can be found in \cite{GK02, KL03}

In order to find the equation of motion of the scalar let us begin with the relation
(\ref{eq:Rev2.19}) multiplied by a sigma matrix  $(\s^a)^{\g\a}$ and differentiate
it with $\nabla_\b$,
\be \eqalign{
(\s^{\un a})^{\g\a}\nabla_{\un a} \left(\nabla_\b\chi_\a\right)
&=~  i \, \frac{1}{2}(\s^{\un a \, \un b})_\a{}^\g\nabla_\b\left(T_{\un a
\, \un b}{}^\a ~-~ 2\g^\prime {\cal R}^{\un k \, \un l}{}_{\un a \,
\un b} \, T_{kl}{}^\a\right)   \cr
&~~~~+~ (\s^{\un a})^{\g\a}\left[\nabla_{\un a} ,\nabla_\b\right]\chi_\a
~~~.}  \label{eq:Rev2.21}
\ee

Notice that the LHS contains the spacetime derivatives of both $\displaystyle{
(\s^{\un b})_{\b \a}\nabla_{\un b} \Phi}$ and $\displaystyle{(\s^{[3]})_{\b\a} \\
L_{[3]}}$, while the RHS can be computed using at most three-half
dimensional results recalled above. Therefore, one obtains the
equation of motion of the scalar from (\ref{eq:Rev2.21}) by taking
the trace $\d_\g{}^\b$
\be
16 \nabla^{\un a}\nabla_{\un a}  \Phi
\ =\ 4 {\cal R}~-~ 8\g^\prime {\cal R}^{\un k \, \un l \, \un a \, \un b }
{\cal R}_{\un k \, \un l \, \un a \, \un b} ~+~ \textrm{fermions} ~~~.
\label{eq:Rev2.22}
\ee
Moreover, the same relation (\ref{eq:Rev2.21}), if multiplied by
$(\s_{\un e \, \un f})_\g{}^\b$, yields
\be
\nabla^{\un a}L_{{\un a \, \un e \, \un f} }\ =\ -4 L_{\un a \, \un e \, \un f }
\nabla^{\un a}  \Phi ~+~ \textrm{fermions} ~~~.  \label{eq:Rev2.23}
\ee

The remaining independent part of (\ref{eq:Rev2.21}) can be
projected out if one multiplies it by $(\s_{efgh})_\g{}^\b$. The
obtained relation together with the Bianchi identity for the
torsion with only vectorial indices gives
\be
\nabla_{[\un
e}L_{\un f \, \un g \, \un h]}\ =\ -3L_{ [{\un e} \,{\un
f}}{}^{\un a} L_{{\un g} \, {\un h}] \, {\un a}} ~-~
\frac{3}{2}\g^\prime {\cal R}_{{\un k} \, {\un l}\,[{\un e} \,
{\un f}}{\cal R}^{{\un k} \, {\un l}}{}_{ {\un g} \, {\un h}]}~+~
\textrm{fermions} ~~~. \label{eq:Rev2.24}
\ee

Notice that (\ref{eq:Rev2.23}) and (\ref{eq:Rev2.24}) suggest that
the object $L_{\un a \un b \un c}$ might be {\em {either}} related
to the field strengths of a two-form {\em {or}} dual field
strength of a six-form depending on which of these two equations
is interpreted as the Bianchi identity and which is as the
equation of motion.

For example, assuming that (\ref{eq:Rev2.23}) gives the equation
of motion for a two-form gauge field, then (\ref{eq:Rev2.24}) must
correspond to its Bianchi identity.  Searching for a closed
three-form in the geometry, in which the field strengths of this
two-form can be identified, one might want to use the identity
satisfied by a Lorentz Chern-Simons three-form $Q$\footnote{ The
second two indices on the Riemann curvature tensor may be thought
of as the Lie \newline $~~~\,~~$ algebraic ``group'' indices for
SO(1,9). } \be \nabla_{[\un e}Q_{\un f \, \un g \, \un h]} ~-~
\frac{3}{2}T_{[\un e \, \un f}{}^{\un a} \, Q_{\un g \, \un h] \,
\un a} ~=~ -\, \frac{3}{2} {\cal R}_{[\un e \, \un f| \un k \, \un
l} \,{\cal R}_{|\un g \, \un h]}{}^{\un k  \, \un l} +
\textrm{fermions}  ~~~
\label{eq:Rev2.25}
\ee
in order to ``absorb" the curvature squared term in the RHS of
(\ref{eq:Rev2.24}) .

Observe that the structure of the equations (\ref{eq:Rev2.24}) and
(\ref{eq:Rev2.25}) is almost the same, with the only difference
that in the RHS the role of the ``group'' indices and ``form''
indices of the curvature are exchanged with respect to one
another. Since the curvature is defined by a connection with
torsion, it is not symmetric with respect to the exchange of its
pairs of indices. Therefore, $(L-\g^\prime Q)_{\un a \un b \un c}$
cannot be equal exactly to the vectorial component of a closed
three-form, but their difference is an object which serves as a
link between the two curvature squared expressions we have in
(\ref{eq:Rev2.24}) and (\ref{eq:Rev2.25}).  This object (called
``$Y_{
\un a \un b \un c}$'' in the next chapter) does exist as was first demonstrated
in \cite{10DCSform2}.  After it has been properly identified, we can use
$Y_{\un a \un b \un c}$ to show
\be
\nabla_{[\un e}(L \,-\, \g^\prime Q \,-\,  \g^\prime Y)_{\un f \, \un g \, \un h]}
-\frac{3}{2}T_{[\un e \, \un f}{}^{\un a} (L \,-\, \g^\prime Q \,-\, \g^\prime Y)_{
\un g \, \un h] \, \un a}
\ =\ \textrm{fermions}
\label{eq:Rev2.26}
\ee
at first order in $\g^\prime$. This is the relation, which shows that (at least
modulo fermionic contributions) $(L-\g^\prime Q-\g^\prime Y)_{\un a \un b
\un c}$ can be identified as the vectorial component of a closed three-form.

Conversely assuming that (\ref{eq:Rev2.24}) gives the equation of motion
for a two-form gauge field, then (\ref{eq:Rev2.23}) must correspond to its
Bianchi identity in the dual theory.  This theory is slightly easier to construct
because although it contains the first order corrections superstring corrections,
it does not require a dual Chern-Simons term for its consistency.

Finally, the Ricci tensor and the scalar curvature can be derived
from (\ref{eq:Rev2.20}) using the dimension three-half results
\bea
\frac{1}{2}{\cal R}^{(\un d \, \un c)}&=&2\nabla^{(\un d} \nabla^{\un c)} \Phi
~+~ 2 \, \g^\prime \, {\cal R}^{\un k \, \un l \, \un d \, \un b}\, {\cal R}_{\un k \, \un l}{}^{
\un c}{}_{\un b} ~+~\textrm{fermions} ~~~,
\label{eq:Rev2.27} \\
{\cal R}&=&-16\nabla^{\un a} \Phi  \nabla_{\un a} \Phi+\frac{2}{3}L^{{\un a \,
\un b \, \un c} }L_{{\un a \, \un b \, \un c} }+3\g^\prime \, {\cal R}^{{\un k \, \un l \,
\un a \, \un b} } \, {\cal R}_{{\un k \, \un l \, \un a \, \un b} } ~+~ \textrm{fermions}
~~~.
\label{eq:Rev2.28}
\eea

Throughout our discussion up to this point, we were working directly with
the superfields of 10D, $\cal N$ $=~1$ superspace supergravity.  So all
equations were superspace equations.   For the rest of this paper, we will
set {\em {all}} fermions to zero.  We will utilize the same symbols to denote
the various quantities however.  We establish the following notation for
the purely bosonic equations found from the superspace Bianchi identities,
\bea
\hat{\ce}_\Phi&\doteq&4\nabla^{\un a} \nabla_{\un a} \Phi ~-~ {\cal R} ~+~
2 \g^\prime \, {\cal R}^{{\un k \, \un l \, \un a \, \un b} } \,
{\cal R}_{{\un k \, \un l \, \un a \, \un b} }   ~~~,
\label{eq:Rev2.29} \\
\hat{\ce}_{B_{\un e \un f}}&\doteq&\nabla_{\un a} \left( \, e^{4 \Phi}L^{\un a \, \un e \,
\un f} \, \right)  ~~~,
\label{eq:Rev2.30} \\
\hat{\ce}_{\widetilde{B}^{\un e \, \un f \, \un g \, \un h}}&\doteq&\nabla_{[\un e}
( \, L \,-\, \g^\prime Q \,-\, \g^\prime Y)_{\un f \, \un g \, \un h]}
-\frac{3}{2}T_{[\un e \, \un f}{}^a (L \,-\, \g^\prime Q-\g^\prime Y)_{\un g \, \un h]
\un a}  ~~~,~~
 \label{eq:Rev2.31}   \\
\hat{\ce}_{\eta_{\un d \un c}}&\doteq&\frac{1}{2}{\cal R}^{(\un d \, \un c)}~-~ 2\nabla^{
(\un d} \nabla^{\un c)} \Phi ~-~ 2\g^\prime {\cal R}^{\un k \, \un l \, \un d \un b}
\, {\cal R}_{\un k \, \un l}{}^{\un c}{}_{\un b}
~~~,
\label{eq:Rev2.32}  \\
\hat{\ce}_{\eta}&\doteq&{\cal R} ~+~ 16\nabla^{\un a} \Phi \nabla_{\un a}\Phi
~-~ \frac{2}{3} \,L^{\un a \, \un b \, \un c}L_{\un a \, \un b \, \un
c} ~-~ 3\g^\prime {\cal R}^{{\un k \, \un l \, \un a \, \un b} } \, {\cal R}_{{\un k
\, \un l \, \un a \, \un b} }  ~~~.
\label{eq:Rev2.33}
\eea
In order for the superspace Bianchi identities to be satisfied all
of the $\Hat \ce$-quantities are required to vanish.  The question
with which we shall wrestle for the rest of this paper is, ``Does
there exist a component level action whose variations lead to
equations of motion that  are compatible with (\ref{eq:Rev2.29}) -
(\ref{eq:Rev2.33})?''  This same action must also contain a field
such that either (\ref{eq:Rev2.30}) or (\ref{eq:Rev2.31}) can be
interpreted as a Bianchi identity.

\section{Bosonic Terms of a Component Action for Two-form Formulation }

~~~~ The non-vanishing components of the modified 3-form field strength to this order
can be written as (below we have used a slightly different set of conventions
from  \cite{10DCSform2} as discussed in an appendix)
\bea
H_{\a\b \, \un c}&=&i \frac{1}{2}(\s_{\un c})_{\a\b}
\,+\, i \,4 \,\g^\prime(\s_{\un a})_{\a\b} \, G^{\un a \, \un e \, \un f}G_{\un c \,
\un e \, \un f}  ~~~,    \label{eq:Rev3.32} \\
H_{\a \un b \, \un c}&=&i \, 2 \,\g^\prime\left[ ~ (\s_{[ \un b|})_{\a\b}T_{\un e
\un f}{}^\b \,-\, 2(\s_ {[\un e ]})_{\a\b}T_{\un f \, [\un b|}{}^\b ~\right] \, G_{\un c ]}{}^{
\un e \, \un f} ~~~,    \label{eq:Rev3.33} \\
H_{\un a \, \un b \, \un c}&=&G_{\un a \, \un b \, \un c} \,+\, \g^\prime
Q_{\un a \, \un b \, \un c} ~~~.
 \label{eq:Rev3.34}
\eea
 In the
limit where $\g'$ $=~0$ these equations correspond to the superspace geometry
in a string-frame description of the pure supergravity theory.   As was pointed
out some time ago \cite{10DCSfrngFrame}, the field {\em {independence}} of
the leading term in the $ G_{\a \b \un c} $ component of the 3-form field strength
is indicative of this.

The quantity $L_{\un a \, \un b \, \un c}$ in this formulation is defined by,
\bea
L_{\un a \, \un b \, \un c}&\doteq&G_{\un a \, \un b \, \un c} \,+\, \g^\prime Q_{
\un a \, \un b \, \un c} \,+\, \g' \, Y_{\un a \, \un b \, \un c} ~+~ \co((\g^{\prime)^2})
~~~,
\label{eq:Rev3.35}
\eea
where $G_{\un a \, \un b \, \un c}$  is the supercovariantized field strength of a two-form,
$Q_{\un a \, \un b \, \un c}$ is the Lorentz Chern-Simons form and
\bea
Y_{\un a \, \un b \, \un c}&\doteq&- \, \left( ~ {\cal R}^{\un e \, \un k}{}_{[\un a
\, \un b |} \,+\, {\cal R}_{[\un a \, \un b|}{}^{\un e \, \un k} \,+\,
\frac{8}{3} \,G_{\un d}{}^{\un e}{}_{[\un a|}G^{\un d \, \un k}{}_{|\un b|} ~ \right)
\, G_{|\un c] \, \un e\,  \un k} ~~~.
\label{eq:Rev3.36}
\eea
This quantity, (which to our knowledge first appeared in \cite{10DCSform2}) has a
remarkable property.  It is a straightforward calculation to show
\be
\nabla_{[\un e}Y_{\un f \, \un g \, \un h]} ~-~ \frac{3}{2}T_{[\un e
\, \un f}{}^{\un a} \, Y_{\un g \, \un h] \, \un a} ~=~ -\, \frac{3}{2}
\left(~ {\cal R}_{\un k \, \un l[ \un e \, \un f}{\cal R}^{\un k \, \un l}
{}_{\un g \, \un h]} ~-~ {\cal R}_{[\un e \, \un f| \un k \, \un l} \,{\cal R}_{|\un g \,
\un h]}{}^{\un k  \, \un l}   \, \,\right) ~+~ \co(\g^\prime) ~~.
\label{eq:Rev3.37}
\ee

By keeping terms {\em {only}} up to first order in $\g'$ we find
that a Lagrangian density of the form \be \cl\ =\ {\rm e}^{-1}
e^{4\Phi} \,\left[~ {\cal R}({\o}) ~+~16\, ({\rm e}^{\un a} \Phi)
\, ({\rm e}_{\un a} \Phi) ~-~ \frac{1}{3} \, L^{\un a \, \un b \,
\un c} \, L_{\un a \, \un b \, \un c} ~+~ \g^\prime {\rm
{tr}}({\cal R}^{\un a \, \un b}{\cal R}_{\un a \, \un b})\right]
~~~, \label{eq:Rev3.38} \ee where $\o$ is the torsion-less spin
connection, is compatible with the set of equations of motion
(\ref{eq:Rev2.27}), (\ref{eq:Rev2.28}),(\ref{eq:Rev2.30}),
(\ref{eq:Rev2.31}) and Bianchi identity (\ref{eq:Rev2.29}). If we
expand the penultimate term to first order in $\g'$ we find \be
\eqalign{ \cl\ =\ {\rm e}^{-1} e^{4\Phi} \, {\Big [}~ &{\cal
R}({\o}) ~+~16\, ({\rm e}^{\un a} \Phi) \, ({\rm e}_{\un a} \Phi)
~-~ \frac{1}{3} \, G^{\un a \, \un b \, \un c} \, {\Big (} \,
G_{\un a \, \un b \, \un c} \,+\, 2 \g' \,Q_{\un a \, \un b \, \un
c}   \, {\Big)}
 {~~~~~~~~~~~~~}  \cr
&~-~ \frac{2}{3} \, \g^\prime  \, G^{\un a \, \un b \, \un c} \,
Y_{\un a \, \un b \, \un c} ~+~ \g^\prime {\rm {tr}}({\cal R}^{\un a \, \un b}{\cal R}_{\un a
\, \un b}) ~ {\Big ]} ~~~.}
\label{eq:Rev3.39}
\ee
It is easily seen that the action to first order in $\g'$ when written using the
$Y$ variable takes a simple and elegant form.

Variation of this Lagrangian with respect to the dilaton gives
\be
\d_\Phi \cl\ \sim\ -4{\rm e}^{-1} e^{4 \Phi} \, \left[ ~\ce_g ~+~ 2 \ce_\Phi
~ \right]  \, \d \Phi
~~~.
\label{eq:Rev3.40}
\ee
where $\ce_g $ and $\ce_\Phi $ are given by (\ref{eq:Rev2.33}) and (\ref{eq:Rev2.29}).

The variation with respect to the antisymmetric tensor at first seems very
complicated due to the fact that its field strength appears in the  Lorentz
connection. However, one can write it simply as
\be
\d_B\cl\ =\ {\rm e}^{-1} e^{4 \Phi}\left(-\frac{2}{3}L^{\un a \, \un b \, \un c}\,
\d L_{\un a \, \un b \, \un c} ~+~ \g^\prime\d{\rm tr}\,({\cal R}^{\un a \, \un b} \,
{\cal R}_{\un a \, \un b})\right) ~~~.
\label{eq:Rev3.41}
\ee
Replacing now  (\ref{eq:Rev3.35}) into the first term, we obtain the form
\be \eqalign{
\d_B\cl\ \sim\ &2\ce_{B_{\un a \un b}}\d B_{\un a \, \un b}
-\frac{2}{3}{\rm e}^{-1}\g^\prime \, e^{4 \Phi}L^{\un a \, \un b \, \un c}\d_L
\left(Q+Y\right)_{\un a \, \un b \, \un c}   \cr
&+~{\rm e}^{-1}\, e^{4 \Phi}\g^\prime
\d_L{\rm tr}\,\left(~{\cal R}^{\un a \, \un b} \, {\cal R}_{\un a \, \un b}  ~\right) ~~~.}
\label{eq:Rev3.41b}
\ee
The last terms in fact form a combination of variations which can
be expressed in terms of zero order equations of motion for
arbitrary variations of the entire object $L_{\un a \, \un b \,
\un c}$. This is shown in appendix B, where this combination is
denoted symbolically by $f(\ce)$.  Therefore, the variation of the
Lagrangian with respect to the antisymmetric tensor is
\be
\d_B\cl\ \sim\ 2\ce_{B_{\un a \un b}}\d B_{\un a \, \un b}
~+~ \g^\prime \, f(\ce)
\label{eq:Rev3.42}
\ee
with
\be \eqalign{
f(\ce)&\sim~4\ce_{B_{\un k \un l}}\F_{\un k}{}^{\un a \, \un b}\d L_{\un l \,
\un a \, \un b} \cr
&~~~~+~8\left[\, e^{4\Phi}\nabla^{\un a}\left(\, e^{-4\Phi}\hat{\ce}_{
B_{\un b \un c}}\right) ~+~ \left(\, e^{4\Phi}\hat{\ce}_{\eta_{\un a \un
k}} \,-\, \hat{\ce}_{B_{\un a \un k}}\right) L_{\un k}{}^{\un b \, \un
c}\right]\d L_{\un a \, \un  b \, \un c}
\cr
&~~~~-~\frac{2}{3} \, \frac{1}{4!} \, e^{4\Phi} \, \ce_{\widetilde{B}^{\un a \un b
\un c \un d}} \d_L\left(\ce_{\widetilde{B}_{\un a \, \un b \, \un c \, \un d
}}\right) ~+~\co(\g^\prime)~~~.  }
\label{eq:Rev3.43}
\ee
\section{Bosonic Terms of a Component Action for Six-form Formulation }

~~~~Retaining the same the current $A_{\un a \un b \un c}$ specified by
(\ref{eq:Rev2.12}) we can introduce a seven-form $N$ \cite{NewD10SG}
satisfying an appropriate Bianchi identity.  At the component level similar
considerations have been carried out for the six-form formulation \cite{BergdROO2}.
One of the remarkable things about this formulation is that in order to describe lowest
order perturbative contributions to the effective does {\em {not}} require a Chern-Simons
like modification to the seven-form field strength.
\bea
N_{\a\b [5]}&=&i \, \frac{1}{2} \, e^{4\Phi} \,(\s_{[5]})_{\a\b}~~~,
\label{eq:Rev4.44} \\
N_{\a [6]}&=&-\frac{1}{4!}\e_{[6][4]} \, e^{4\Phi} \,
(\s^{[4]})_\a{}^\b\chi_\b~~~,
 \label{eq:Rev4.45}\\
N_{[7]}&=&\frac{1}{3!} \, e^{4\Phi}  \, \left(L^{[3]}-\frac{13i
\,}{8}\chi\s^{[3]}\chi\right) \e_{[3][7]} ~~~. \label{eq:Rev4.46}
\eea
In particular, it is the equation (\ref{eq:Rev2.29}) which
insures that the purely vectorial component of the $N$ Bianchi
identity is satisfied.  Equations (\ref{eq:Rev2.29}),
(\ref{eq:Rev2.30}), (\ref{eq:Rev2.32}) and (\ref{eq:Rev2.33})
contain the bosonic equations of motion for the component fields
of the dual theory.  Notice that in this case (\ref{eq:Rev2.29})
identifies $ L_{\un a \un b \un c}$ as the following function of
the component fields of the dual theory \be L_{\un a \, \un b \,
\un c}\ =\ -\frac{1}{7!} \, \e_{\un a \, \un b \, \un c[7]}  \,
e^{-4\Phi}  \, N^{[7]} ~~~. \label{eq:Rev4.47} \ee upon setting
the fermions to zero.  In the following we show that the
Lagrangian density
\be
\eqalign{ \cl_d\ =\ {\rm e}^{-1} \,
e^{4\Phi} \, {\Big [} ~ &{\cal R}({\o}) ~+~ 16 \, ({\rm e}^{\un
a}\Phi)\, ({\rm e}_{\un a}\Phi) ~+~ \frac{1}{3}(~L \,-\,
\g^\prime(Q+Y))^2_{\un a \, \un b\, \un c} \cr &~+~ \g^\prime \,
{\rm tr}({\cal R}^{\un a \, \un b}{\cal R}_{\un a \, \un b}) ~
{\Big ]}  ~~~,   } \label{eq:Rev4.48}
\ee
is compatible with the
set of equations of motion and Bianchi identity.  Since our
results are only valid to first order in $\g'$ it follows that
(\ref{eq:Rev4.48}) should be more properly written as
\be
\eqalign{ \cl_d\ =\ {\rm e}^{-1} \, e^{4\Phi} \, {\Big [} ~ &{\cal
R}({\o}) ~+~ 16 \, ({\rm e}^{\un a}\Phi)\, ({\rm e}_{\un a}\Phi)
~+~ \frac{1}{3} \,L^{\un a \, \un b\, \un c} \, L_{\un a \, \un
b\, \un c} \cr &~-~ \frac{2}{3}\, \g^\prime  \, L^{\un a \, \un
b\, \un c} \,  Q_{\un a \, \un b\, \un c} ~-~ \frac 23\, \g^\prime
\,L^{\un a \, \un b\, \un c} \, Y_{\un a \, \un b\, \un c} ~+~
\g^\prime \, {\rm tr}({\cal R}^{\un a \, \un b}{\cal R}_{\un a \,
\un b}) ~ {\Big ]}  ~~~,   } \label{eq:Rev4.48b}
\ee
and in this expression $L$ is replaced by the expression in
(\ref{eq:Rev4.47}).  When this is done two points are made
obvious.  Firstly, this action is not in the string-frame
formulation. This follows in particular since the object $L_{\un a
\un b \un c}$ depends on the dilaton through (\ref{eq:Rev4.47}).
>From the superspace point of view this was already obvious due to
the field dependence exhibited by (\ref{eq:Rev4.44}).  A
string-frame formulation of the dual theory does exist after
additional field redefinitions are applied to (\ref{eq:Rev4.48})
and (\ref{eq:Rev4.48b}).

Secondarily, the Chern-Simons term does not actually appear in this action.  One
can perform an integration-by-part on the first term on the second line of (\ref{eq:Rev4.48b})
and this leads
to a term
\be \eqalign{
 L^{\un a \, \un b\, \un c} \,  Q_{\un a \, \un b\,
\un c} ~\propto~ \e^{{\un a}_1  \, \cdots  \, {\un a}_6 \, {\un b}_1 \, {\un b}_2 \,
{\un c}_1 \, {\un c}_2 \, } \, M_{{\un a}_1 \, \cdots \,  {\un a}_6} \,
{\rm tr}({\cal R}_{{\un b}_1 \, {\un b}_2}{\cal R}_{{\un c}_1 \, {\un c}_2} \, )
 ~~~,   }
\label{eq:Rev4.48c}
\ee
which can be seen to be precisely the term required by the dual Green-Schwarz
mechanism for anomaly cancellation first given in \cite{NewD10SG}.  Notice
the change of sign of the $L$-squared term in (\ref{eq:Rev4.48}) and
(\ref{eq:Rev4.48b}) compared to (\ref{eq:Rev3.38}) and (\ref{eq:Rev3.39}).  This
is the usual sign-flip seen between theories connected by Poincar\' e duality.

Indeed, now even the variation with respect to the dilaton becomes
complicated since $L_{\un a \un b \un c}$ appears in the connection.
However, just marking the variation and using $\d L=-4L\d\Phi$ only in the
most obvious terms, one ends again with the combination of variations
$f(\ce)$ near the terms of the equation for the dilaton in the theory with
two-form (\ref{eq:Rev3.43}),
\bea
\d_\Phi \cl_d&\sim&-4\, {e}^{-1} \,e^{4\Phi} \, \left[ ~ \ce_g
~+~ 2\ce_{\Phi}  ~\right]\, \d\Phi ~+~ \g^\prime f(\ce)
~~~.
\label{eq:Rev4.49}
\eea
The variation with respect to the six-form $M$ is computed in the same manner.
As a ``miracle" the combination $f(\ce)$ surprisingly appears again and one
simply obtains,
\be
\d_M\cl_d\ \sim\ -\frac{2}{3}\frac{1}{4!6!} \e^{\un a \, \un b \, \un c \, \un d \,
[6]}\hat{\ce}_{\widetilde
{B}_{\un a  \un b  \un c  \un d }}\d M_{[6]}+\g^\prime f(\ce) ~~~.
\label{eq:Rev4.50}
\ee
So the final conclusion is that in the dual theory, the component action
in (\ref{eq:Rev4.48b}) is compatible with the equations of motion derived
from superspace for the dual theory.

\section{Comparison with a Component Level Investigation}

~~~~Next, let us study the relationship of the Lagrangian (\ref{eq:Rev3.39})
with the component Lagrangian in \cite{BergdROO}. A quick look to the component
Lagrangian in \cite{BergdROO} convinces us that using just rescalings
of the fields it can be written in the form
\be \eqalign{
\hat{\cl}\ =\ {\rm e}^{-1} \, e^{4 \Phi}{\Big [} ~&{\cal R}(\o) ~+~16\, ({\rm e}^{\un a} \Phi) \,
({\rm e}_{\un a} \Phi) ~-~ \frac{1}{3} \, G^{\un a \, \un b \, \un c} \,  (\,
G_{\un a \, \un b \, \un c} \,+\, 2 \g' \, Q_{\un a \\, \un b \, \un c} \,)
{~~~~~~~~~~~~~}  \cr
&~+~ \g^\prime {\rm {tr}}({\Hat {\cal R}}^{\un a \, \un b}\,
{\Hat {\cal R}}_{\un a \, \un b}) ~ {\Big ]} ~~~,}
\label{eq:Rev5.51}
\ee
where hatted objects are defined using a Lorentz connection
$\Hat{\O}$, which may differ from ours by its torsion. In order to
compare this to our Lagrangian (\ref{eq:Rev3.39}), let us write
the difference as
\be \eqalign{
{\rm e}^{-1} \,e^{-4\Phi} {\Big(} ~ \cl-\hat{\cl} {\Big)}&=~-\frac{2}{3}\g^\prime
G^{\un a \, \un b \, \un c} \, \left(Q-\hat{Q}\right)_{\un a \, \un b \, \un c} ~-~ \frac{2}{3}
\g^\prime G^{\un a \, \un b \, \un c} \, Y_{\un a \, \un b \, \un c} \cr
&~~~~~+~ \g^\prime \, {\rm tr}\left(~{\cal R}_{\un a \, \un b} \, {\cal R}^{\un a \, \un b}\,-\,
{\Hat {\cal R}}_{\un a \, \un b} \, {\Hat {\cal R}}^{\un a \, \un b} ~\right) ~~~. }
\label{eq:Rev5.53}
\ee
Observe that the difference is in fact a $GY$ term. The question
is whether this additional term can be removed by field
redefinitions.

First of all, notice, that only redefinitions at zero order of the
Lorentz connection can affect this difference at first order. For
example, let us redistribute the torsion in the connection using a
real parameter $k$ in the simplest way,
\bea
{ \O}_{\un a \, \un b \, \un c}&=& \o_{\un a \, \un b \, \un c}~-~
L_{\un a \, \un b \, \un c}\ =\ {\Hat \O}_{\un a \, \un b \,
\un c} ~+~ \chi_{\un a \, \un b \, \un c}  ~~~,
\label{eq:Rev5.55}   \\
{\Hat \O}_{\un a \, \un b \, \un c}&=&\o_{\un a \, \un b \, \un c}
~-~ (1-k)L_{\un a \, \un b \, \un c}
~~~,  \label{eq:Rev5.56}\\
\chi_{\un a \, \un b \, \un c}&=&-kL_{\un a \, \un b \, \un c} ~~~.
\label{eq:Rev5.57}
\eea
This can be seen as a shift in the connection of type
(\ref{Appx}), which is frequently used to find conventional
constraints in supergravity. For $k=0$ in fact there is ``no
redefinition", for $k=1$ the new connection $\Hat \O
\,=\,
\o$ is torsionfree, while for $k=2$ the sign of the torsion flips.

How does this shift in the connection affect the form of the Lagrangian?
One computes the changes in the Chern-Simons term and the curvature
squared term using (\ref{Appx3}) and respectively (\ref{Appx2})
\bea
\frac{2}{3} \, G^{\un a \, \un b \, \un c} \, {\Big(} \, Q \,- \, \Hat{Q}
 \, {\Big)}_{\un a \, \un b \, \un c}&\sim& -4k\left[~ {\cal R}_{\un a \, \un b \, \un c \,
\un d} \,+\, 2k\left(1-\frac{k}{3}\right) \, G_{\un a \, \un c}{}^{\un k} \, G_{\un b \,
\un d \, \un k} \right] \, G^{\un a \, \un b \, \un l}G^{\un c \, \un d}{}_{\un l}
\nonumber\\
&&-4k\, e^{-4 \Phi}\hat{\ce}_{B_{\un a \un b}} \, {\Hat \O}_{\un b \, \un e \,
\un k} \, G_{\un a}{}^{\un e \, \un k} ~+~ \co(\g^\prime) ~~~,
\label{eq:Rev5.58}  \\[2mm]
-\frac{2}{3} G^{\un a \, \un b \, \un c} \, Y_{\un a \, \un b \, \un c}&=&8 \,
 \left[ ~ {\cal R}_{\un a \, \un b \, \un c \, \un d} \,+\, \frac{4}{3} \, G_{\un a \, \un
c}{}^{\un k} \, G_{\un b \, \un d \, \un k} \right] \, G^{\un a \, \un b \, \un l}
\, G^{\un c \, \un d}{}_{\un l}  ~~~,
\label{eq:Rev5.59}   \\[2mm]
{\rm tr}\left( ~{\cal R}_{\un a \, \un b} \, {\cal R}^{\un a \, \un b} \,-\, \Hat{{\cal R}}_{\un a \, \un b}
\, \Hat{{\cal R}}^{\un a \, \un b}\right)
&\sim&2k^2 \,(k-2)^2 \,\left[~ \left(G_{\un a \, \un b}{}^{\un k} \, G_{\un c \, \un d
 \, \un k}  \,-\, G_{\un a \, \un c}{}^{\un k} \,G_{\un b \, \un d \, \un k} \right) ~
\right] \,G^{\un a \, \un b \, \un l} \, G^{\un c \, \un d}{}_{\un l}
\nonumber\\
&&-4k(k-2)\left[\hat{\ce}_{\eta_{\un k \un l}}G_{\un l}{}^{\un c \, \un d \,}
+\nabla^{\un k}(\,e^{-4 \Phi}\hat{\ce}_{B_{\un c \, \un d \,}})\right]
\, G_{\un k \, \un c \, \un d \,} \nonumber \\
&&+~2k(k-2)\, {\cal R}_{\un a \, \un b \,\un c \, \un d \,} G^{\un a \,
\un b \, \un k} \, G^{\un c \, \un d \,}{}_{\un k} ~+~ \co(\g^\prime) ~~~,
\label{eq:Rev5.60}
\eea
and finally we find
\bea
e^{-4 \Phi}\left(~\cl \,-\, \Hat{\cl} ~\right)
&\sim&2(k-2)^2\g^\prime {\cal R}_{\un a \, \un b \,\un c \, \un d \,}
\, G^{\un a \, \un b \, \un k} \, G^{\un c \, \un d \,}{}_{\un k}
\nonumber \\
&&+(k-2)^2\g^\prime
\left[2k^2G_{\un a \, \un b \,}{}^{\un k} \, G_{\un c \, \un d \, \un k} \,+\,
 \frac{2}{3}(k+4)G_{\un a \, \un c}{}^{\un k} \, G_{\un b \, \un d \, \un k}
\right]G^{\un a \, \un b \, \un l}\, G^{\un c \, \un d \,}{}_{\un l}
\nonumber\\
&&-4k(k-2)\g^\prime \left[~ \Hat{\ce}_{\eta_{\un c \un l}}G_{\un l}{}^{
\un k \, \un d} \,+\, \nabla^{\un k}(\, e^{-4 \Phi}\hat{\ce}_{B_{\un
c \,\un d \,}}) ~\right] \, G_{\un k \, \un c \, \un d \,} \nonumber \\
&&-4k\g^\prime \,
e^{-4 \Phi}\, \Hat{\ce}_{B_{\un c \, \un d \,}}\, { \Hat \O}_{\un c
\, \un e \, \un f} \, G_{\un d}{}^{\un e \, \un f} ~~~.
\label{eq:Rev5.61}
\eea
Observe, that for $k=0$, indeed, the difference is equal to the
$GY$ term, while for $k=2$, the difference is a term proportional
to the equation of motion for the antisymmetric tensor at zero
order:
\bea
\cl \,-\, \Hat{\cl}
&\sim&-8\g^\prime \,
 \Hat{\ce}_{B_{\un c \, \un d \,}}\, { \Hat \O}_{\un
c
\, \un e \, \un f} \, G_{\un d}{}^{\un e \, \un f} ~~~.
\eea

At first sight it seems that the change of sign of the torsion in
the Lorentz connection just exchanges the $GY$ term to another
``unwanted" one. However, correction terms which are proportional
to equations of motion can be absorbed by field redefinitions
involving the perturbation parameter and therefore $\cl$ and
$\hat{\cl}$ are equivalent.

Indeed, let us consider the expression
\be
S[\phi]+\gamma^\prime\int dx^n
\frac{\delta S}{\delta\phi}\cf(\phi)\,,
\ee
with $S[\phi]$ an action for the fields $\phi$, $\frac{\delta
S}{\delta\phi}=0$ the equations of motion for the fields $\phi$,
$\cf(\phi)$ an arbitrary function of the fields $\phi$ and
$\gamma^\prime$ an infinitesimal parameter. Now consider the field
redefinitions
\be
\phi^\prime\ =\ \phi+\gamma^\prime \cf(\phi),
\ee
and develop $S[\phi^\prime]$ around $\phi$ using that
$\gamma^\prime$ is infinitesimal. Then one obtains
\be
S[\phi^\prime]\ =\ S[\phi]+\gamma^\prime\int dx^n
\frac{\delta S}{\delta\phi}\cf(\phi)
+{\cal{O}}({\gamma^\prime}^{2}).
\ee

As a conclusion, we have proven here that the bosonic Lagrangian
(\ref{eq:Rev3.39}), based on the superspace geometry proposed in
\cite{10DCSform2} is equivalent to the component-level first-order
corrected supergravity Lagrangian of \cite{BergdROO}.

\section{Conclusion }

~~~~ With this present work, we have re-engaged in a discussion
that began almost twenty years ago. We hope that this has
presented in the clearest possible terms the proposal given in
\cite{10DCSform2}.
In particular, we gave the bosonic Lagrangian corresponding to the
superspace geometry proposed in \cite{10DCSform2} and we showed
that this Lagrangian is equivalent to the gravity part (the YM
coupling constant is set to zero) of the first-order corrected
anomaly-free supergravity Lagrangian.

The issue of duality in the framework of superspace geometry is
discussed and the dual theory is also presented.

Aside from issues connected with the controversy over the form of
the lowest order terms (deformations) in the slope-parameter
expansion of the supergeometry, the technique in this paper of
using superspace to generate equations of motion, taking the limit
of vanishing fermions and then integrating the resulting bosonic
equations to derive an action insures that the bosonic limit
reached in this manner is consistent with supersymmetry. We
believe this paper marks one of the first times this set of steps
has been applied to a supergravity theory.

A discussion of these results and their relation to some of the
``conventional wisdom'' on this topic (based on \cite{IT}) will be
treated in a separate publication \cite{Pap2}.

 \vspace{.1in}
 \begin{center}
 \parbox{4in}{{\it ``It's no exaggeration to say the undecideds could
    go one ${\,}$ way or another.''}\,\,-\,\,George H. W. Bush}
 \end{center}

 \vspace{.2in}

 \noindent

 {\bf Acknowledgements}\\[.1in] \indent

SJG wishes to acknowledge the hospitality and assistance of
Maliwatch, the organizational committee for the 2004 Mali
Symposium on Applied Sciences held in Bamako, Mali during August 1
- 6, 2004. Additional acknowledgements go to the organizers of the
Einstein Celebration held at the Aspen Institute of Physics during
Aug. 9 - 11, 2004. AK wishes to acknowledge the hospitality of the
University of Maryland and in particular of the Center for String
and Particle Theory. \newline $~~$ \newline

\noindent
{\Large{\bf Appendix A:  Definitions \& Conventions}}

~~~~ The basic tool we use is ten dimensional chiral superspace with
structure group $SO(1,9)$. Definitions and properties
(such as multiplication table and Fierz identities) of ten
dimensional chiral sigma matrices
we adopted here can be found in \cite{PACKG}.
Given the super frame $E^\ca=(E^{\un a},\ E^\a)$,
conventions for superforms and Leibniz rule for the exterior derivative are
\bea
\o&=&\frac{1}{p!}E^{\ca_1}...E^{\ca_p}\, \o_{\ca_p...\ca_1}\,,\\
{\rm d}(\o_p\o_q)&=&\o_p({\rm d}\o_q)+(-)^q({\rm d}\o_p)\o_q\,.
\eea

Representation matrices acting on the tangent space are blockdiagonal,
\be
X\ =\ \left(\begin{array}{cc}X_b{}^a&0\\
0&X_\b{}^\a\end{array}\right),
\ee
and the vectorial and spinorial representations are related by the
two-index sigma matrix,
\be
X_\a{}^\b\ =\ \frac{1}{4}(\s^{\un a \, \un b})_\a{}^\b X_{\un a \, \un b},\qquad
X_{\un a \, \un b}\ =\ -\frac{1}{8}(\s_{\un a \, \un b})_\a{}^\b X_\b{}^\a.
\ee
As soon as the action of the structure group is fixed,
\be
\d E\ =\ \b E X,
\ee
the covariant derivative
\be
\nabla E\ =\ {\rm d} E \,+\, \a E\O
\ee
can be defined using the Lorentz connection $\O$ with transformation law
\be
\d\O\ =\ -\b\left( {\rm d} X\,+\, \a X \cdot\O\right)\,.
\ee

The torsion $T$, the curvature ${\cal R}$ and field strengths $F_p$ of an
abelian $(p-1)$-form are defined by \be \nabla E\ =\ \g T,\qquad
{\cal R}\ =\ {\rm d}\O+\a\O\O,\qquad F_{p}\ =\ {\rm d} A_{p-1}, \ee and
they satisfy the following Bianchi identities \be \g \nabla T\ =\
\a E{\cal R}\label{BI_T},\qquad \nabla {\cal R}\ =\ 0,\qquad {\rm d} F_{p}\ =\
0. \ee The curvature in particular appears in the double covariant
derivative of covariant vectors \be \nabla \nabla u\ =\ \a u{\cal R}.
\label{AppXx} \ee Dragon's theorem states that in supergravity the
Bianchi identity for the torsion together with (\ref{AppXx})
implies that the Bianchi identity for the curvature is
automatically satisfied.

The Chern-Simons form
\be
Q\ =\ \, {\rm tr}\left(\O {\cal R}-\frac{\a}{3}\O\O\O\right)
\label{Appa}
\ee
satisfies
\be
{\rm d} Q\ =\ {\rm tr}({\cal R}{\cal R})\,.
\ee

Finally, let us consider a redefinition
\be
\O\ =\ \hat{\O}+\chi
\label{Appx}
\ee
of the connection. This shift in the connection affects the torsion,
the curvature and the Chern-Simons form in the following way:
\bea
\g (T-\hat{T})&=&\a E\chi  ~~~, \label{Appx1} \\
{\cal R}-\hat{{\cal R}}&=&\nabla \chi-\a\chi\chi ~~~, \label{Appx2}\\
Q-\hat{Q}&=&\, {\rm tr}\left(2{\cal R}\chi-\chi \nabla \chi+\frac{2\a}{3}\chi\chi\chi
+{\rm d}(\O\chi)\right)
~~~. ~~~, \label{Appx3}
\eea

Let us display the above relations in terms of form-components.
First of all, (\ref{AppXx}) gives the algebra of covariant
derivatives acting on covariant vectors \be
\left(\nabla_\cc,\nabla_\cb\right)u^\ca\ =\ -\g
T_{\cc\cb}{}^\cf\nabla_\cf u^\ca +\a {\cal R}_{\cc\cb\cf}{}^\ca
u^\cf\,. \label{AppYY} \ee The Bianchi identities become \bea
\g\nabla_{(\nabla }T_{\cc\cb)}{}^\ca +\g^2T_{(\nabla\cc|}{}^\cf
T_{\cf|\cb)}{}^\ca
-\a {\cal R}_{(\nabla\cc\cb)}{}^\ca&=&0\\
\nabla_{(\ca_1}F_{\ca_2...\ca_{p+1})}
+\g \frac{p}{2}T_{(\ca_1\ca_2|}{}^\cf F_{\cf|\ca_3...\ca_{p+1})}&=&0\,.
\eea

The components of the Chern-Simons form are
\be
Q_{\ca\cb\cc}\ =\ \, {\rm tr}\left(\frac{1}{2}\O_{(\ca}{\cal R}_{\cb\cc)}
+\frac{\a}{3}\O_{(\ca}\O_\cb\O_{\cc)}\right),
\ee
while the redefinitions take the form
\be \eqalign{ {~~~~~}
\g(T-\hat{T})_{\cc\cb}{}^\ca&=~\a\chi_{(\cc\cb)}{}^\ca ~~~,  \cr
({\cal R}-\hat{{\cal R}})_{\cb\ca}&=~ \nabla_{(\cb}\chi_{\ca)}+\g T_{\cb\ca}{}^\cf\chi_\cf
+\a\chi_{(\cb}\chi_{\ca)}  ~~~,  \cr
(Q-\hat{Q})_{\cc\cb\ca}&=~  {\rm tr} {\Big[}~ {\cal R}_{(\cc\cb}\chi_{\ca)}
-\chi_{(\cc}\left(\nabla_\cb\chi_{\ca)} +\frac{\g}{2}T_{\cb\ca)}{}^\cf\chi_\cf
+\frac{2\a}{3}\chi_\cb\chi_{\ca)}\right)  ~~~ , \cr
&-\nabla_{(\cc}(\O_\cb\chi_{\ca)}
-\frac{\g}{2} \O_{(\cf}\chi_{(\ca)}T_{\cc\cb)}{}^\cf {\Big]}   ~~~.}
\label{Appw}
\ee

The conventions of Wess and Bagger correspond to the choice $\a=1$,
$\g=1$, while
the conventions in \cite{10DCSform2} correspond to  $\a=-1$, $\g=-1$. Also, the
Chern-Simons term denoted by $X$ in \cite{10DCSform2} is $X=-Q$.

The graviton and gravitino is identified in the super frame
$E^\ca=(E^{\un a},\, E^\a)$, \be E^{\un a} \doubar\ =\ {\rm d}
x^{\un m} {\rm e}_{\un m}{}^{\un a},\qquad E^\a\doubar\ =\
\frac{1}{2}{\rm d} x^{\un m} \psi_{\un m}{}^\a. \ee The torsion,
$T\ =\ -\nabla E$, satisfies the Bianchi identity \be \nabla T\ =\
E{\cal R}. \label{Appq} \ee The two-form gauge potential of the pure 10
dimensional supergravity multiplet is identified in a two-form on
the superspace \be B\doubar\ =\ \frac{1}{2}{\rm d} x^{\un m}{\rm
d} x^{\un n} B_{\un n \, \un m}. \ee Its fieldstrengths $G={\rm d}
B$ satisfies the Bianchi identity \be {\rm d} G\ =\ 0.
\label{Appqq} \ee The Green-Schwarz mechanism teaches us that in
order to deal with anomaly free supergravity the field strength of
the antisymmetric tensor has to be accompanied by both the
Yang-Mills and gravitational Chern-Simons terms. Here we consider
only the gravitational part. \be Q\ =\ {\rm
tr}({\cal R}\O+\frac{1}{3}\O\O\O),\qquad {\rm d} Q\ =\ {\rm tr}({\cal R}{\cal R}). \ee
Therefore, it is convenient in general to define a new object on
superspace, \be H\ \doteq\ G\ +\ \g^\prime Q, \ee and consider the
Bianchi identity satisfied by this three-form $H$, \be {\rm d} H\
= \g^\prime {\rm tr}({\cal R}{\cal R}). \label{AppP} \ee The six-form gauge
potential of the dual pure 10 dimensional supergravity multiplet
is identified in a six-form on the superspace \be M\doubar\ =\
\frac{1}{6!}{\rm d} x^{{\un m}_1}...{\rm d} x^{{\un m}_6} M_{{\un
m}_6\, \dots \, {\un m}_1}. \ee Its fieldstrengths $N={\rm d} B$
satisfies the Bianchi identity \be {\rm d} N\ =\ 0. \label{AppPp}
\ee

\newpage
\noindent
{\Large{\bf Appendix B:  Variations}}

~~~~ For arbitrary variation of the connection $\d \O$ the curvature squared terms
and the Chern-Simons form $Q$ change according to
\bea
\d\, {\rm tr}\left({\cal R}^{\un a \un b}\, {\cal R}_{\un a \un b}\right)& =& -4\, {\rm tr}
\left[(\nabla_{\un a} \, {\cal R}^{\un a \un b})\, \d\O_{\un b}\right] ~+~ 4\, \partial_{
\un m} {\Big(} \,  {\rm e}_{\un a}{}^{\un m} \, {\rm tr}({\cal R}^{\un a \un b}\, \d\O_{\un
b}) \,  {\Big)} ~~~, \\
\d Q&=&\textrm{tr}\left[~2{\cal R}\d\O\,+\, {\rm d}(\O\d\O) ~\right] ~~~.
\label{AppA}
\eea
The scalar curvature transforms also:
\bea
\d {\cal R}&=&{\rm e}_{\un a}{}^{\un m} \,{\rm e}_{\un b}{}^{\un n} \, \d {\cal R}_{\un m
\, \un n}{}^{\un a \, \un b}    \\
&=&2{\rm e}_{\un a}{}^{\un m}\, \partial_{\un m} (\d\O_{\un b}{}^{\un a \, \un b})
~-~ T_{\un a\, \un b}{}^{\un c} \, \d\O_{\un c}{}^{\un a \, \un b}  ~~~.
\eea
In the case where $\d\O_{{\un a \, \un b \, \un c}}=\frac{1}{2}\d T_{{\un a \, \un b \, \un
c}}$ with totally antisymmetric torsion this yields $\d {\cal R}=-\d(\frac{1}{4}T_{{\un a \, \un
b \, \un c}}T^{{\un a \, \un b \, \un c}})$. In particular this  implies also that the
combination $ {\cal R}+\frac{1}{4}T_{{\un a \, \un b \, \un c}}T^{{\un a \, \un b \, \un c}}$ is
independent of a redefinition (\ref{Appx})  provided that $\chi$ is totally antisymmetric.

Using the above formulae one may compute the following variations
with respect to
an object $L_{\un a \un b \un c }$ appearing in the Lorentz connection as
\be
\O_{\un a \, \un b \, \un c}\ =\ \o_{\un a \, \un b \, \un c} ~-~ L_{\un a \, \un b \, \un c},
\ee
with $\o$ the torsion free spin connection:
\bea
{\rm e}^{-1}\, e^{4\Phi} \d_L{\rm tr}\left(\, {\cal R}^{\un a \, \un b} \, {\cal R}_{\un a \, \un b}
\, \right)
&\sim&-4{\rm e}^{-1}\nabla^{\un a}(\, e^{4\Phi}{\cal R}_{\un a \, \un b \, \un c \, \un d}) \,
\d L^{\un b \, \un c \, \un d} \nonumber  \\
&&~+~ 4{\rm e}^{-1}\, e^{4\Phi}{\cal R}_{\un a \, \un b \, \un c \, \un d} \, L^{
\un a \, \un b}{}_{\un k} \, \d L^{\un k \un c \, \un d}  \nonumber \\
&&~+~ \co(\g^\prime) ~~~,
\label{AppVR}\\
-\frac{2}{3}{\rm e}^{-1} \, e^{4\Phi}L^{\un a \, \un b \, \un c}\d_L Q_{\un a \, \un
b \, \un c} &\sim&4\ce_{B_{\un k \un l}}\O_{\un k}{}^{\un a \, \un b}\, \d L_{\un l
\, \un a \, \un b} ~-~ 4{\rm e}^{-1} \, e^{4\Phi}\, {\cal R}_{\un a \, \un b \, \un c \, \un d}
\, L^{\un a \, \un b}{}_{\un k}\d L^{\un k \, \un c \, \un d}
 \nonumber  \\
&&~+~ \co(\g^\prime)\\
-\frac{2}{3}{\rm e}^{-1} \, e^{4\Phi}L^{\un a \, \un b \, \un
c}\d_L Y_{\un a \, \un b \, \un c} &=& 4\, e^{4\Phi}({\cal R}_{{\un a
\, \un b}{\un c \, \un d}}+{\cal R}_{{\un c \, \un d}{\un a \, \un
b}})L_{\un k}{}^{{\un a \, \un b}}  \, \d L^{\un k \,\un c \, \un d}
\nonumber\\
&&-\frac{2}{3} \, e^{4\Phi} L_{\un a \, \un b}{}^{\un k} \, L_{{\un c \, \un d} \, \un k}
\d_L \left(\ce_{\widetilde{B}_{\un a  \un b \un c \un d}   }    \right)
~+~ \co(\g^\prime).
\eea
However, the first term in the variation (\ref{AppVR}) may be recast in the form
\bea
-4\, {\Big[} \, \nabla^{\un a}(\, e^{4\Phi}{\cal R}_{\un a \, \un b \, \un c\, \un d})
\, {\Big]} \, \d L^{\un b \, \un c\, \un d} &\sim&
-\, 4\, e^{4\Phi}({\cal R}_{{\un a \, \un b}{\un c \, \un d}}
\,+\, {\cal R}_{{\un c \, \un d}{\un a \, \un b}})L_{\un k}{}^{{\un a
\, \un b}}  \, \d L^{\un k \, {\un c \, \un d}}       \nonumber\\
&&+\, 8\left[\, e^{4\Phi}\nabla^{\un a}\left(\, e^{-4\Phi}
\hat{\ce}_{B_{{\un b \, \un c}}}\right) \, \right] \,
\d L_{\un a \un b \, \un c}          \nonumber\\
&&+\, 8\left[\, \, \left(\,
e^{4\Phi}\hat{\ce}_{\eta_{\un a \un k}}
\,-\, \hat{\ce}_{B_{\un a \un k}}\right) L_{\un k}{}^{{
\un b\,\un c}}\right]\d L_{\un a \un b \, \un c}     \nonumber\\
&&+\, \frac{2}{3} \, e^{4\Phi}\left[ ~L_{{\un a \, \un b}}{}^{\un k}
\,  L_{\un  c \, \un d \, \un k} \,-\,\frac{1}{4!}\ce_{\widetilde{B
}^{\un a \un b \un c \un d}} ~\right] \,
\d_L\left(\ce_{\widetilde{B}_{\un a \un b \un c  \un d}}\right)
\nonumber\\
&&
+\, \co(\g^\prime).
\eea
Now observe, that the sum of the variations written above
is expressed as a combination of the equations we derived
from superspace geometry. We denote this combination of variations
simbolically by $f(\ce)$:
\bea
f(\ce)& \doteq& {\rm e}^{-1}\, e^{4\Phi} \d_L{\rm tr}\left({\cal R}^{\un a \, \un b}
\, {\cal R}_{\un a \, \un b}\right) ~-~
\frac{2}{3}{\rm e}^{-1}\, e^{4\Phi}\, L^{\un a \, \un b \, \un c}  \, \d_L \left(
~Q \,+\, Y~ \right)_{\un a \, \un b \, \un c}  ~~~,
\label{comb}\\
f(\ce)&\sim&4\ce_{B_{\un k \un l}}\O_{\un k}{}^{\un a \, \un b}\d L_{
\un l \, \un a \, \un b}    \nonumber\\
&&+~8\left[\, e^{4\Phi}\nabla^{\un a}\left(\, e^{-4\Phi}\hat{\ce}_{B_{\un b
\un c}}\right) \,+\, \left(\, e^{4\Phi}\hat{\ce}_{\eta_{\un a \un k}} \,-\,
\hat{\ce}_{B_{\un a \un k}}\right) \, L_{\un k}{}^{\un b \, \un c}\right]
\, \d L_{\un a \, \un b \, \un c}\nonumber\\
&&-~\frac{2}{3}\, e^{4\Phi} \frac{1}{4!}\ce_{\widetilde{B}^{\un a \, \un b \, \un c\,
\un d}} \d_L\left(\ce_{\widetilde{B}_{\un a \un b \un c \un d}} ~ \right) \\
&&+~\co(\g^\prime).
\eea
Therefore the superspace equations imply the vanishing of the above
combination for an arbitrary variation of the object $L_{\un a \un b \un c}$.
In particular, this is valid at zero order in $\g^\prime$ both for the anomaly free
supergravity and for its dual.

\end{document}